\documentclass[aps,pre,showpacs,amsmath,amssymb,floatfix,
superscriptaddress,nofootinbib]{revtex4}

\usepackage{xcolor,wasysym}

\usepackage{graphicx}
\usepackage{bm}
\usepackage{eucal}
\usepackage{mathrsfs}
\graphicspath{{figures/}}

\newcommand {\be} {\begin{equation}}
\newcommand {\ee} {\end{equation}}
\newcommand {\Be}{\begin{eqnarray*}}
\newcommand {\Ee} {\end{eqnarray*}}
\newcommand {\bey} {\begin{eqnarray}}
\newcommand {\eey} {\end{eqnarray}}
\newcommand{\bit}{\begin{itemize}}      
\newcommand{\eit}{\end{itemize}}
\newcommand{\bfl}{\begin{flusleft}}
\newcommand{\efl}{\end{flusleft}}
\newcommand{\bfr}{\begin{flushright}}
\newcommand{\bc}{\begin{center}}
\newcommand{\ec}{\end{center}}
\newcommand{\ben}{\begin{enumerate}}    
\newcommand{\een}{\end{enumerate}}

\newcommand{\comment}[1]{}

\usepackage[caption=false,font=small]{subfig}
\usepackage{placeins}
\usepackage[draft]{fixme}
\newcommand{\w}{\ensuremath{W}}
\newcommand{\tweakedSim}{{\raise.17ex\hbox{$\scriptstyle\sim$}}}

\begin{document}

\title{Sisyphus Effect in Pulse Coupled Excitatory Neural Networks \\ 
with Spike-Timing Dependent Plasticity}
\date{\today}

\author{Kaare Mikkelsen}
\affiliation{Dept. of Physics and Astronomy, University of Aarhus,
Ny Munkegade, Building 1520 - DK-8000 Aarhus C, Denmark}

\author{Alberto Imparato}
\affiliation{Dept. of Physics and Astronomy, University of Aarhus,
Ny Munkegade, Building 1520 - DK-8000 Aarhus C, Denmark}

\author{Alessandro Torcini}
\affiliation{CNR - Consiglio Nazionale delle Ricerche - Istituto dei Sistemi 
Complessi, via Madonna del Piano 10, I-50019 Sesto Fiorentino, Italy}
\affiliation{INFN Sez. Firenze, via Sansone, 1 - I-50019 Sesto Fiorentino, Italy}
\affiliation{Dept. of Physics and Astronomy, University of Aarhus,
Ny Munkegade, Building 1520 - DK-8000 Aarhus C, Denmark}

\begin{abstract}
The collective dynamics of excitatory pulse coupled neural networks with spike 
timing dependent plasticity (STDP) is  studied. Depending on the model parameters 
stationary states characterized by High or Low Synchronization can be observed. In particular, 
at  the transition between these two regimes, persistent irregular low
frequency oscillations between strongly and weakly synchronized states are
observable, which can be identified as infraslow oscillations with 
frequencies $\simeq 0.02 - 0.03$ Hz.
Their emergence can be explained in terms of the Sisyphus Effect, a mechanism
caused by a continuous feedback between the evolution of the  coherent 
population activity and of the average synaptic weight.
Due to this effect, the synaptic weights have oscillating equilibrium values, 
which prevents the neuronal population from relaxing into a stationary macroscopic state.
\end{abstract}


\maketitle
  
\section{Introduction}

Fluctuating spontaneous neural activity has been observed
in several areas of the brain: ranging from the cortex to the
hippocampus, from the thalamus to the basal ganglia
and cerebellum~\cite{harris,van2011slow}. In particular, low-frequency
fluctuations (LFFs), in the range 0.5-1 Hz, have been observed
in the cortical local field potential (LFP) during sleep as well as during quiet 
wakefulness~\cite{steriade2007brain,okun2010}. In the hippocampus, 
this kind of irregular oscillations between states, characterized by higher and lower
levels of synchrony, have been revealed during slow-wave sleep and
related to the process of memory consolidation in the neocortex~\cite{buzsaki1989}.
Infraslow oscillations (ISOs), corresponding to frequencies $\simeq 0.02 - 0.2$ Hz,
have been identified in humans using high-density fullband
electroencephalography during 
the execution of somatosensory detection tasks and during sleep~\cite{monto2008,vanhatalo2004infraslow,van2011slow}.
Furthermore, these infraslow oscillations have been associated
to a cyclic modulation of cortical excitability,
possibly related to the aggravation of epileptic activity during sleep~\cite{vanhatalo2004infraslow}.

Synaptic plasticity is a fundamental ingredient of neuronal activity,
being involved in the transfer of information and in its processing
at the neuronal and population level. On the one hand, 
during the sleep-wake cycle, distinct oscillatory patterns
organize the activity of neuronal populations, 
modulating synaptic plasticity~\cite{romcy2009synaptic}. 
On the other hand, synaptic plasticity has been identified as one of the 
fundamental mechanisms at the origin of multistable states in neural circuits
\cite{Tass2006Longterm, Maistrenko2007Multistability,Lubenov2008,ciszak2011,mongillo2012}.
In particular, spike-timing dependent plasticity (STDP) represents an important, 
experimentally measured, mechanisms controlling the strength of the synapse connecting a pre-synaptic 
to a post-synaptic neuron. 
STDP is a temporally asymmetric form of Hebbian learning, based on the causal relationship
between the spikes emitted at pre- and postsynaptic neurons
~\cite{Markram1997Regulation,Bi1998Synaptic, debanne1998, Song2000Competitive,
Froemke2002Spiketimingdependent, Wang2005Coactivation}.
A spike emitted in the pre-synaptic cell, within a certain time interval
(learning window), before an emission in the
postsynaptic cell triggers long-term potentiation (LTP), whereas the 
reversed temporal order in spike emission results in long-term depression (LTD)~\cite{STDP}.
Learning windows for LTD and LTP are asymmetric, as clearly 
shown experimentally \cite{Bi2001Synaptic, Bi2002Temporal, FD}.

It has been shown, both experimentally \cite{Nowotny2003Enhancement} 
and theoretically \cite{Zhigulin2003Robustness,Zhigulin2004Important}, that STDP
influences the collective behavior of a neural population, generally
leading to an increase in the degree of synchronization. 
However, in the presence of propagation delays 
STDP can instead provide a negative feedback mechanism contrasting
highly synchronized network activity~\cite{Lubenov2008}.
The presence of noise in this latter case can
lead to the emergence of states at the boundary between 
randomness and synchrony, while in oscillatory neural
networks the desynchronizing effect due to noise
is counteracted by the STDP action~\cite{Chen2010Meanfield,Chen2011,popovych2013self}.
On a general ground the asymmetry in the learning windows 
seems a prerequisite for the emergence of coexisting 
states with different degree of 
synchronization~\cite{Tass2006Longterm,Maistrenko2007Multistability,Pfister2010STDP}.
  
In this paper, we study the  {\it Sisyphus Effect} (SE), a 
deterministic mechanism recently
introduced to explain the spontaneous emergence of irregular
oscillations in the neural population activity in presence of STDP~\cite{mikkelsen2013}.
In particular, we include STDP in the renowned neural network
model developed in~\cite{abbott_vanvreeswik} by Abbott and van Vreeswijk. This model is composed of
leaky integrate-and-fire (LIF) neurons and the synaptic interactions are purely excitatory 
and mediated by $\alpha$-pulses \cite{Rall1967Dendritic}. In absence of plasticity, 
the network activity is asynchronous for slow synapses (and/or large synaptic weights)
and partially synchronized for sufficiently 
fast synapses (and/or weak coupling)~\cite{Vreeswijk1994,hansel1995,vanVreeswijk1996Partial}.
Similar to Sisyphus, who was bound in Tartarus for the eternity 
to roll a boulder uphill just to watch it rolling back down again,
STDP leads the asynchronous system towards a synchronized state
by modifying the synaptic weights, however as soon as the synchronous
regime is achieved the weights are attracted back towards their starting
values and the system desynchronizes. Thus STDP should repeat its action again and again, 
and this leads to endless oscillations in the synchronization level of the population activity.

The paper is organized as follows, Sect. II is devoted to the introduction of the model,
of the integration scheme as well as of the indicators employed to characterize the degree 
of synchronization of the neural population and its collective activity. The collective
dynamics in plastic and non-plastic neural networks is described in Sect. III with particular
emphasis on the emergence of LFFs. Sect. IV  reports a characterization of the synchronization
oscillations in the neural activity in terms of the evolution of an order parameter on 
an effective free energy landscape. The results of simulations, performed by maintaining a
constant average synaptic weight, are analyzed in Sect. V, while Sect. VI is
devoted to the mean-field analysis of the synaptic weights evolution. The SE is illustrated
in Sect. VII for different synaptic parameters. Finally a brief summary and a synthetic
discussion of the reported results can be found in Sect. VIII. In Appendix A the stationary distributions
of the synaptic weights are displayed for different parameters, while Appendix B 
reports the transformations required to convert the variables and parameters entering in the model
from adimensional units to physical ones.

\section{Model and Indicators}

\subsection{Neural Network Model}

\noindent We study a fully coupled network composed of $N$ Leaky Integrate-and-Fire (LIF)
neurons, whose membrane potentials $V_i(t) \in [V_{r}:V_{th}]$ are ruled by the following equation:
\begin{equation}\label{eq:V1}
\dot{V}_{i}(t)= a-V_{i}(t)+I_i(t)\, \quad\quad i=1,\cdots, N
\\ \quad ;
\end{equation}
where $I_i$ is the synaptic current due to the inputs received from the rest of the network,
$a$ is the intrinsic excitability of the neuron which can be due to nonspecific 
background currents arising from distant brain areas or to external DC current terms,
in particular we assume that the neuron is suprathreshold, i.e. $a > 1$.
Whenever the neuron $i$ reaches the 
threshold $V_{th} \equiv 1$, a pulse $p_\alpha(t)$ is instantaneously transmitted to all the other
neurons and its membrane potential is reset to $V_{r} \equiv 0$. The synaptic current
can be written as $I_i(t) = g E_i(t)$, with $g  >  0$ representing
the excitatory {\it homogeneous} synaptic strength, while the field $E_i(t)$ is
given by the linear superposition of all the pulses $p_\alpha(t)$ received
by neuron $i$ in the past. The formal expression of $E_i(t)$ reads as
\begin{equation}\label{eq:E0}
E_i(t)= \frac{1}{N-1} \sum_{n|t_n < t} w_{ij}(t_n) \Theta(t-t_n) p_\alpha(t-t_n)
\quad ,
\end{equation}
where $N-1$ is the number of pre-synaptic neurons, since autapses have
been avoided, $\Theta(t)$ is the Heaviside function and $w_{ij}$ 
represents the synaptic weight associated to a directional link
connecting the pre-synaptic neuron $j$ to the post-synaptic one $i$ at the time of spike emission, $t_n$.
The scaling of the field $E_i$ with the number of synaptic inputs reported 
in (\ref{eq:E0}) is intended to mimic the homeostatic synaptic scaling 
experimentally observed for excitatory neurons~\cite{Turrigiano2008}.
 
Following van Vreeswijk~\cite{vanVreeswijk1996Partial}, we assume 
$\alpha$-function shape for the pulses, i.e. $p_\alpha(t)=\alpha^2 t \exp(-\alpha t)$.
The time evolution of the field $E_i(t)$ is thus ruled by 
the following second order differential equation:
\begin{equation}
\label{eq:E}
\ddot E_i(t) +2\alpha\dot E_i(t)+\alpha^2 E_i(t)= 
\frac{\alpha^2}{N-1}\sum_{n|t_n< t} w_{ij}(t_n) \delta(t-t_n) \ .
\end{equation}

For a fully coupled network, in absence of plasticity,
the weights $w_{ij}$ appearing in Eq. (\ref{eq:E0}) and (\ref{eq:E})
are all equal to one (apart from the autaptic terms which are set to zero)
and the fields $E_i$ are all identical, therefore the neurons
are driven by a common field $E$. In presence of plasticity we assume that the synaptic weights
evolve in time according to the STDP rule with soft bounds, namely
\begin{equation}\label{eq:dynW}
\dot{w}_{ij}(t)= p [w_{max}-w_{ij}(t)] A_j S_i - d w_{ij}(t) B_i S_j 
\quad ,
\end{equation}
where $d$ ($p$) is the potentiation (depression) amplitude, and $S_k =
\sum_{n|t_n<t} \delta(t-t_n) $ represents the time series of spikes emitted by
neuron $k$ until time $t$. The presence of the bound implies that $ 0 \le
w_{ij} \le w_{max}$.

The variables $A_j$ can be thought of as concentrations of glutamate
bound to the post-synaptic receptors, or as the fraction of open 
N-methyl-D-aspartate (NMDA) receptors; while $B_i$ it is usually 
associated to the concentration of calcium entering the cell due 
to a back-propagating action potential~\cite{STDP}.

In particular, we have implemented the {\it nearest neighbor} version of the STDP
rule~\cite{STDP,Chen2010Meanfield}, where the synapses have memory just of the last
emitted spike. In this case the time evolution of the $A_j$ and $B_i$ variables 
is given by
\begin{equation}\label{eq:prepost}
\tau_{+} \dot{A_j} = - A_j +(1-A_j) S_j \qquad
, \qquad
\tau_{-} \dot{B_i} = - B_i +(1-B_i) S_i 
\\ \quad ,
\end{equation}
where $\tau_{+}$ ($\tau_{-}$) are the time scales at 
which post- (pre-) synaptic spikes will cause potentiation 
(depression) of the synapse. As pointed out by Izhikevich and Desai~\cite{Izhikevich2003BCM} 
the nearest neighbor implementation of the STDP rule is consistent with the classical long-term
potentiation and depression as represented in the form of the Bienenstock-Cooper-Munro 
synapse~\cite{BCM}.

Therefore, in the case of a post-synaptic (pre-synaptic)
spike, emitted by neuron $i$ ($j$) at time
$t$, the weight $w_{ij}$ is potentiated (depressed) as
\begin{equation}
 w_{ij}(t^+) = w_{ij}(t^-)+\Gamma_{ij}(t)
\label{wrule}
\end{equation}
with
\begin{equation}
\Gamma_{ij}(t) = \left\{\begin{array}{rcl}
p[w_{M}-w_{ij}(t^-)]{\rm e}^{-\frac{\delta_{ij}}{\tau_{+}}}
& \mbox{if} & \delta_{ij} > 0\\
\label{post}
& &\\
- d \, w_{ij}(t^-){\rm e}^{+\frac{\delta_{ij}}{\tau_{-}}}
& \mbox{if} & \delta_{ij} < 0
\end{array}\right.
\end{equation}
where $\delta_{ij} = t - t^{(j)} > 0$ ($\delta_{ij} = t^{(i)} - t < 0$) is the firing
time difference and $t^{(k)}$ the last firing time of neuron $k$.
The resulting distributions of the synaptic weights and their properties of stationarity
are discussed in the Appendix A.

In this paper, we assume that $\tau_{-} > \tau_{+}$, 
as suggested by the experimental data~\cite{Bi1998Synaptic,FD}, 
and in particular we fix $\tau_{-} = 3 \tau_{+} = 0.3$. 
Furthermore, despite that the main part of the reported results refer to $d= p=0.01$ 
({\it Symmetric Case}, SC), we have also examined a more realistic situation where $p > d$
(as suggested by the experiments reported in ~\cite{Bi1998Synaptic,FD}), namely
by considering $p = 2 d = 0.02$ ({\it Asymmetric Case}, AC). If not explicitly
stated the SC will be studied for $\alpha=9$ and the AC for $\alpha=11$.
The studied model is adimensional,
however it can easily be transformed to physical units as shown in Appendix B.

\subsection{Simulation Method}

\noindent Since the plasticity rule depends critically on the precision of the spiking
events, it is necessary to employ an accurate integration scheme to update
the evolution equations. This makes an event-driven algorithm an optimal
choice, because it conjugates high accuracy in the determination of the spike
times with a fast implementation \cite{Chen2011}. In particular, by following 
Olmi et al. \cite{Olmi2010Oscillations} the event-driven map can be written as
\begin{eqnarray}
E_i(n+1) &=& E_i(n) {\rm e}^{-\alpha \tau(n)}+P_i(n)\tau(n) 
{\rm e}^{-\alpha \tau(n)} \nonumber 
\\
P_i(n+1) &=& P_i(n)e^{-\alpha \tau(n)}+w_{im}\frac{\alpha^2}{N-1}  
\\
V_{i}(n+1) &=& V_i(n)e^{-\tau(n)}+a(1-e^{-\tau(n)})+g H_i(n) 
i=1,\dots,N \quad ; \quad V_m(n+1) \equiv 0 \, ;
 \nonumber
\label{ee}
\end{eqnarray}
where $P_i \equiv \alpha E_i+\dot E_i$ is an auxiliary
variable, $m$ is the index of the neuron emitting the $n+1$-th spike and
$\tau(n)= t_{n+1}-t_n$ is the network inter-spike interval.
The explicit expression for the nonlinear function $H_i(n)$ appearing
in (\ref{ee}) is
\begin{equation}
\label{eq:F1}
H_i(n) = \frac{{\rm e}^{-\tau(n)} - e^{-\alpha\tau(n)}}{\alpha-1}
     \left(E_i(n)+\frac{P_i(n)}{\alpha-1} \right) 
- \frac{\tau(n) e^{-\alpha\tau(n)}}{(\alpha-1)} P_i(n) \, ;
\end{equation}
for the parameter values considered in this paper ($g > 0$ and $a > 1$),
$H_i(n) > 0$.

The event driven map reported in (\ref{ee})
gives the explicit evolution of the fields $E_i,P_i$ and membrane
potentials $V_i$ from the time $t_n^+$ immediately following the $n$-th
spike emission to time $t_{n+1}$. However, the evolution equation depends
on the inter-spike interval $\tau(n)$, which can be determined only implicitly
by solving the following equation
\begin{equation}
\label{eq:ti2}
\tau(n)=\ln\left[\frac{a-V_m(n)}{a+gH_m(n)-1}\right] \ .
\end{equation}
 
Together with the evolution of the fields and membrane potentials, also the
weights of the afferent and efferent synapses associated to the firing neuron $m$
should be updated as follows
\begin{gather}
\begin{split}
w_{mj}(n+1) &= w_{mj}(n) + p[w_{max}-w_{mj}(n)]{\rm e}^{-\Delta
t_j/\tau_{+}}
\\
w_{jm}(n+1) &= w_{jm}(n) - dw_{jm}(n){\rm e}^{-\Delta t_j/\tau_{-}}
\qquad  j=1,\dots,N  \quad w_{jj} \equiv 0 
\label{eq:discreteW}
\end{split}
\end{gather}
where $\Delta t_j=t_{n+1} - t^{(j)}$ and $t^{(j)}$ is the last firing time of neuron $j$.
Please notice that the synapses evolution (\ref{eq:discreteW}) is performed
after the map evolution (\ref{ee}),  because we assume
that the plasticity is a slower process than spike generation. 

The implementation of the event driven map (\ref{ee}) involves $3N -1$
variables, since the membrane potential of the firing neuron is exactly zero at
each firing event, thus it does not take part in the dynamics. The evolution
of the synaptic weights involves $N^2 -N$ variables, since autapses
have been excluded. Altogether our dynamical systems has $N^2+2N-1$
degrees of freedom. Our implementation of the dynamical
evolution of the network assumes that only one neuron at a time will
reach the threshold, therefore in the case of exact clustering, without any source of 
disorder, our method will fail. However, we have always verified our assumption to be true.

In the following we report two kinds of simulations: {\it Constrained} (CS)
and {\it Unconstrained} (US). The results of CSs are discussed in Sect. V,
during CSs the synaptic weights are constrained to have an average value 
\begin{equation}
W(t) \equiv \frac{1}{N(N-1)}\sum_{i,j} w_{ij}(t)
\qquad ,
\label{ave_w}
\end{equation}
which remains equal to $W_0$. To achieve this result the weights $w_{ij}$ are let to evolve following their dynamics,
as expressed in Eqs.~(\ref{eq:discreteW}), however at regular time intervals  the weights
are rescaled as $w_{ij}/W(t)\cdot W_0$ in order to maintain their average value constant
~\footnote{For $N=200$ we renormalized 
the synaptic weights each $0.2$ time units to avoid drifts in the synaptic average value. 
All the other results discussed in the article refer
to USs, where no constraint was imposed on the dynamics.}

\subsection{Synchronization Indicator and Local Field Potential}

The degree of synchronization of the neuronal population can be characterized in terms
of the order parameter~\cite{winfree,kuram}
\begin{equation}
R(t)=  \left|\frac{1}{N} \sum_k e^{i\theta_k(t)}\right| \quad ,
\label{kura}
\end{equation}
where
\begin{equation}
 \theta_k(t)=2\pi \frac{(t-t^{(k)}_m)}{(t^{(k)}_{m+1}-t^{(k)}_m)}
\label{angle}
\end{equation}
is the phase of the $k$-th neuron at time $t$ between its
$m$-th and $(m+1)$-th spike emission, occurring at times
$t^{(k)}_m$ and $t^{(k)}_{m+1}$, respectively.
A non-zero $R$ value is an indication of partial
synchronization, perfect synchronization is
achieved for $R=1$, while a vanishingly small $R \sim 1/\sqrt{N}$ is
observable for asynchronous states in finite systems.

To better reveal the dynamics on long time scales, we have 
low-pass filtered $R(t)$ by performing the following convolution integral
\begin{equation}
R_f(t) =  \frac{1}{\tau_F} \int_0^{T_M}  R(t-\xi) {\rm e}^{-\xi/\tau_F}
\quad ,
\label{filter}
\end{equation}
where $\tau_F^{-1}$ represents the cut-off frequency, and $T_M >> \tau_F$ is the
integration window.

The local field potential (LFP) can be defined by following \cite{nunez2006,hauptmann2009,popovych2011} 
as
\begin{equation}
LFP(t) \equiv - I_f(t) =  \frac{1}{\tau_F} \int_0^{T_M}  \frac{1}{N} \sum_{i=1}^N I_i(t-\xi) {\rm e}^{-\xi/\tau_F}
\quad ,
\label{LFP}
\end{equation}
where $I_f(t)$ represents the filtered input synaptic current averaged over the ensemble of all neurons.
We can consider $LFP(t)$ as the local field potential generated by our ensemble of neurons if they
would be all located at the same spatial distance from the recording electrode, the low-pass
filtering action of dendrites and of the extracellular medium are taken into account by performing
the convolution integral reported in (\ref{LFP}) \cite{buzsaki2012}.
To compare with experimental measurements where high (low) activity correspond to a minimum (maximum)
in the LFP, we reversed the sign of the filtered synaptic current in (\ref{LFP})~\cite{harris}.
In the following, we have usually employed $\tau_F=40$ and $T_M = 300$ and we considered
$R(t)$ and $I_i(t)$ sampled at equal time intervals $\delta T =1$.

\section{Non plastic versus plastic collective dynamics}

\noindent In this Section we compare the possible collective
dynamics observable at the macroscopic level 
in the non plastic and plastic networks by 
characterizing the different macroscopic attractors
in terms of their degree of synchronization, described by $R$.

\subsection{Non plastic network}

\begin{figure}[htb]
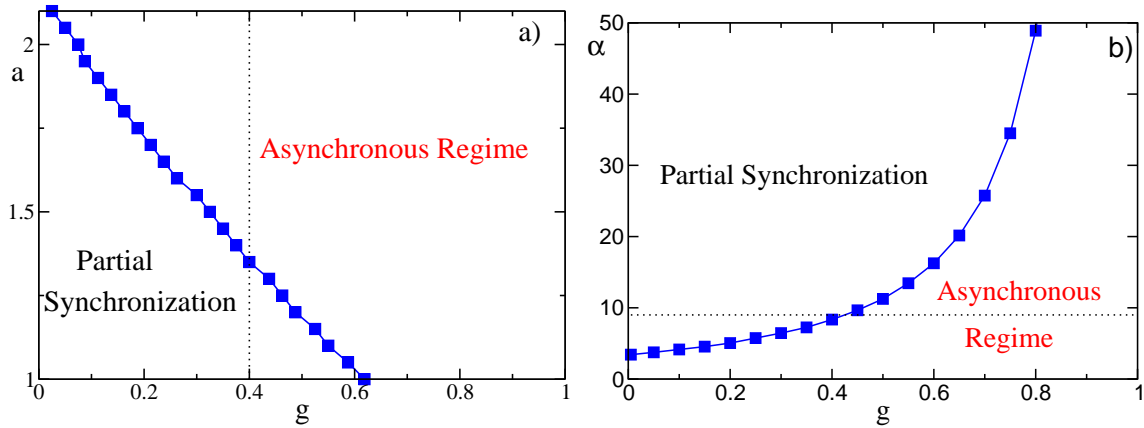

\begin{center}
\includegraphics*[angle=0,width=7.5cm]{fig1a}
\includegraphics*[angle=0,width=7.5cm]{fig1b}
\end{center}
\caption{(color online) 
Phase diagram for the homogeneous model without plasticity,  in (a) we report the phase plane $(g,a)$, while in (b) 
the phases are shown in the $(g,\alpha)$-plane. The blue filled squares indicate the critical values for 
which the asynchronous state (namely, the splay state) becomes unstable; the dotted line indicates $g=0.4$ 
in (a) and $\alpha=9$ in (b). The other parameters are fixed to $\alpha=9$ in (a) and $a=1.3$ in (b) and the 
system size is  $N=100$.
} 
\label{fig:phasediag}
\end{figure}

\noindent In absence of plasticity the dynamics of the model is controlled by three parameters:
namely, the neuronal intrinsic excitability $a$, the synaptic strength $g$
and the inverse pulse width $\alpha$. As shown in Fig. \ref{fig:phasediag}, where the phase
diagrams in the plane $(g,a)$ and $(g,\alpha)$ are reported, only two collective behaviors
can be observed for the non plastic fully connected network: an asynchronous and a partially 
synchronous regime.

In this case the asynchronous regime corresponds to a so-called  {\it splay state} \cite{abbott_vanvreeswik,zillmer2007}:
this is an exact solution for the system which is perfectly asynchronous ($R \equiv 0$). 
Furthermore, this solution is characterized by a constant field $E$ (where the neuron dependence
has been dropped since in this case the fields are all identical) and a periodic evolution of the
membrane potentials. For this solution we are able to perform an exact analytic linear stability 
analysis \cite{zillmer2007}, 
and therefore to determine the stability boundaries of the solution, reported as solid line plus symbols
in Fig. \ref{fig:phasediag}. It is known that whenever the splay state looses its stability 
it gives rise to a partially synchronized regime via a Hopf supercritical bifurcation ~\cite{abbott_vanvreeswik}. 
This regime is characterized, beside a finite value of $R$, by a periodically oscillating macroscopic field $E$
and by a quasi-periodic motion of the membrane potentials at a microscopic level. This regime has
been mainly observed in pulse coupled neural networks \cite{vanVreeswijk1996Partial}, but recently 
partial synchronization has been discovered also in networks of phase oscillators~\cite{rosenblum2007,pikovsky2009}.
and electronic Wien-bridge devices \cite{temirbayev2012experiments} coupled via mean-field nonlinear coupling. 
In this state the single neuron dynamics are quasi-periodic and the field $E$ periodic
with a period which is incommensurate with respect to the single neuron inter-spike interval (ISI)~\cite{Mohanty2006New}.

The transition may be intuitively understood as an interplay between the two time scales present in the model: 
the ISI and the pulse width $1/\alpha$. We observe from 
Fig. \ref{fig:phasediag} (a) and (b) that the asynchronous regime is stable
for small $\alpha$-values and for large $a$- and $g$-values. 
The intrinsic excitability and the (excitatory) synaptic coupling determines the ISI, 
in particular, the ISI duration is a decreasing function of the values of $a$ and $g$.
At sufficiently low $\alpha$, the synaptic current seen
by each neuron is essentially constant, and it induces a stable regular network activity 
corresponding to a periodic firing of successive neurons with a constant population spiking rate.
Whenever the pulse duration drops below a certain value the synaptic input cannot be anymore regarded 
as constant and the state corresponding to a time-invariant network activity becomes unstable.
For $\alpha$-values very large with respect to the ISI, an ``almost'' fully synchronized 
state with $R \simeq 1$ is observed, as expected for excitatory networks where the
transmitted pulses have extremely short rise times, like exponential or 
$\delta$-spikes \cite{Tsodyks1993,Vreeswijk1994,hansel1995}. With reference to the parameter values 
considered in this paper, partial synchronization emerges for $a \le a_c \simeq 1.35 $ for fixed coupling 
$g=0.4$, and for $g \le g_c \simeq 0.4676 $ for fixed pulse width $\alpha=9$,
see Fig. \ref{fig:phasediag}.

\begin{figure}[htb]
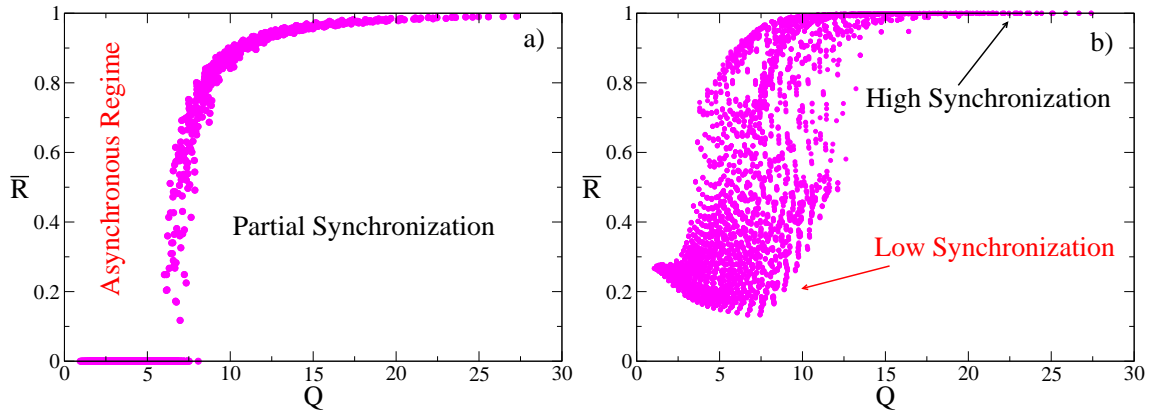

\includegraphics*[angle=0,width=7.5cm]{fig2a}
\includegraphics*[angle=0,width=7.5cm]{fig2b}
\caption{(color online) Time averaged  $\bar R$ as a function of $Q$:
(a) non plastic network; (b) plastic network with $p=d=0.01,\tau_-=3\tau_+=0.3$.
Each plot consists of 1,575 different sets of parameters $(a,g,\alpha)$, 
with $1.1 \leq a \leq 2.5$, $0.1 \leq g \leq 0.9$, $8 \leq \alpha \leq 12$, $N=300$. 
In the non plastic (plastic) case each set of parameters have been simulated 
for 2 (4) different initial conditions and the average performed over a time span
of $10^5$ time units. The measured differences among the various realizations
are smaller than the dimension of the dots.} 
\label{fig:r_ISIalpha}
\end{figure}

In order to characterize the network dynamics it is tempting to introduce
an unique parameter encompassing the two time scales, similarly to what was done
in~\cite{zillmer2007} to analyze the linear stability of the splay state.
This adimensional parameter is the ratio of the two relevant time scales, namely
\begin{equation}
Q =  \frac{\langle ISI \rangle}{1/\alpha} = \alpha \langle ISI \rangle \quad ;
\label{Q}
\end{equation}
where $\langle \cdot \rangle$ denotes an average over the neuronal population
and over the sequence of spikes.
In Fig. \ref{fig:r_ISIalpha} (a) we report the average level of synchronization $\bar R$ for a large
variety of states corresponding to different $(a,g,\alpha)$-triples as a function of $Q$:
as one can notice the data almost collapse onto a universal curve. This indicates
that, to a certain extent, the non plastic network three-dimensional 
phase space $(a,g,\alpha)$ can be described by the single parameter $Q$.
Furthermore, we observe states with $\bar R=0$ for low $Q$ values, namely smaller than
$Q \simeq 8.15$, while a partially synchronized state is observable at large $Q > 6$. Therefore
there is a limited interval $Q \simeq 6 - 8$, where to the same $Q$ value can correspond either
asynchronous or partially synchronized states. However this does 
not necessarily imply a coexistence of the possible attractors,
but simply that the parametrization of the dynamical behaviors 
in terms of an unique parameter $Q$ is not perfect.  
 
\subsection{Plastic network}

\noindent Upon the addition of STDP, the data do not collapse anymore onto a
universal curve, as we would expect since new time scales now
enter in the model microscopic dynamics: namely, the learning time windows. The data reported in 
Fig.~\ref{fig:r_ISIalpha} (b) indicate that at large $Q > 15$ the presence of plasticity essentially 
does not modify the collective behavior already observed in the non plastic network: the system 
remains almost fully synchronized $\bar R \simeq 1$ ({\it High Synchronization} (HS)). 
However, the introduction of plasticity influence the dynamical evolution at smaller $Q$,
for $Q < 2$ the completely asynchronous  state disappears and it is substituted by a regime 
of {\it Low Synchronization } (LS), where $\bar R \simeq 0.3$. Furthermore, in the intermediate $Q$ range 
the neuronal population exhibits a large variability in the level of synchronization,
as measured by the $\bar R$ parameter and this regime will be the main subject of our investigation.
All in all these results suggest that the introduction of STDP favors synchronization in the system.

\begin{figure}[htb]
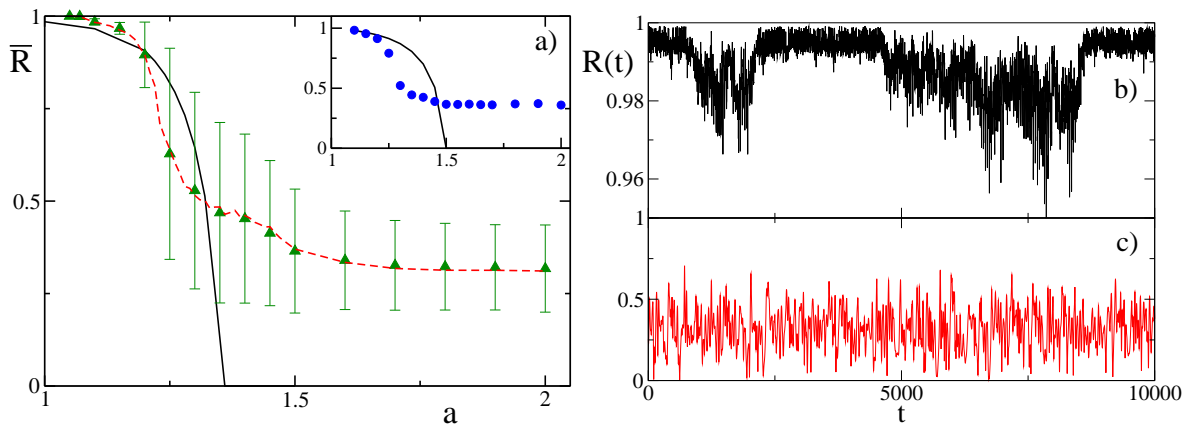

\includegraphics*[angle=0,width=7.5cm]{fig3a}
\includegraphics*[angle=0,width=8cm]{fig3b}
\caption{(color online) 
(a) Average order parameter $\bar R$ as a
function of the intrinsic excitability $a$ for the nonplastic network (black solid line) and
in the presence of STDP for the SC, namely $\alpha=9$ and $d=p=0.01$ (the green  
triangles refer to $N = 200$ and the red dashed lime to $N=500$).
The inset reports the data for  $\alpha=11$ and $N=300$ for the non plastic network
(black solid curve) and with STDP for the AC with $p=2d=0.02$ (blue dots \textcolor{blue}{$\CIRCLE$}).
(b) and (c) Time evolution of $R(t)$ for the SC with $N=500$
for $a=1.09$ and $a=1.70$, respectively.
The data refer to  $g = 0.4$, $\tau_- = 3\tau_+ = 0.3$ and $w_{max}=2$
and the averages have been performed over a time span $\simeq 10^4$,
after discarding a transient $\simeq 10^5$.
} 
\label{fig:r_a}
\end{figure}

In particular, we will focus on the dependence of the macroscopic dynamics on 
the intrinsic excitability $a$. The time averaged order parameter $\bar R$ is reported 
as a function of $a$ in Fig. \ref{fig:r_a} (a) for the non plastic situation as well 
as for the plastic case.

As already stressed, the introduction of STDP destabilizes the asynchronous state that
is now substituted by a Low Synchronization state with $\bar R \simeq 0.32-0.35$ for the
set of parameters considered in the figure. Furthermore, at low excitability $a \simeq 1-1.2$
the systems is almost fully synchronized $\bar R \simeq 1$, but the macroscopic evolution
reveals {\it high frequency fluctuations} (HFFs) (see Fig. \ref{fig:r_a} (b)). For high excitability $a \ge 1.5$
$R(t)$ oscillates quite rapidly around a finite non zero value as shown in Fig. \ref{fig:r_a} (c).
The most part of the reported results refer to equal potentiation 
and depression amplitudes (SC), however these findings are essentially confirmed
also for the more realistic AC, as shown in the inset of Fig. \ref{fig:r_a} (a). 

\begin{figure}[htb]
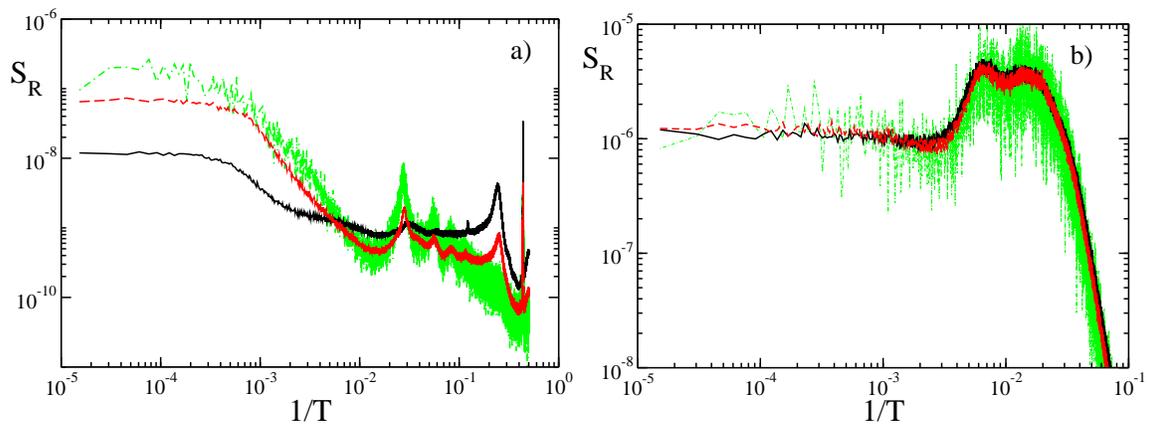

\includegraphics*[angle=0,width=7.5cm]{fig4a}
\includegraphics*[angle=0,width=7.5cm]{fig4b}
\caption{(color online) 
Power spectra of the order parameter for different excitability,
namely $a=1.10 $ (a) and $a=1.90$ (b). The (black) solid line
refers to $N=100$, the (\textcolor{red}{red}) dashed to $N=200$ and the (\textcolor{green}{green}) dot-dashed
to $N=400$ neurons. The data refer to SC for $g = 0.4$, $\alpha=9$, $d=p=0.01$, 
$\tau_- = 3\tau_+ = 0.3$ and $w_{max}=2$ 
and are obtained after discarding a transient $\simeq 10^5$.
The power spectra have been obtained by the time trace of the parameter $R$
sampled at regular intervals $\delta T = 1$ and for a time window
$T_{tot} = 65,536$, the spectra are averaged over 300-600 different time windows 
at $N=100$ and 200 over 5-10 windows for $N=400$.
} 
\label{fig:spectra}
\end{figure}

To better characterize the regimes observable in the SC, we have estimated the power spectra $S_R$ associated to $R(t)$
reported in Fig. \ref{fig:spectra}, the spectrum for $a=1.1$ resembles a Lorentzian with 
subsidiary peaks at low periods (namely $T < 35$). The Lorentzian part of the spectrum can be fitted as
$\simeq 1/(\lambda^2 +T^{-2})$, thus indicating that  
it originates from a Poissonian point process with a relaxation time  $\lambda^{-1} \simeq 1,400 - 2,000$.
Furthermore, the  $S_R$ corresponding to $a=1.9$ reveals two nearby HFF peaks at $T \simeq 70$ and 150.
 
\begin{figure}[htb]
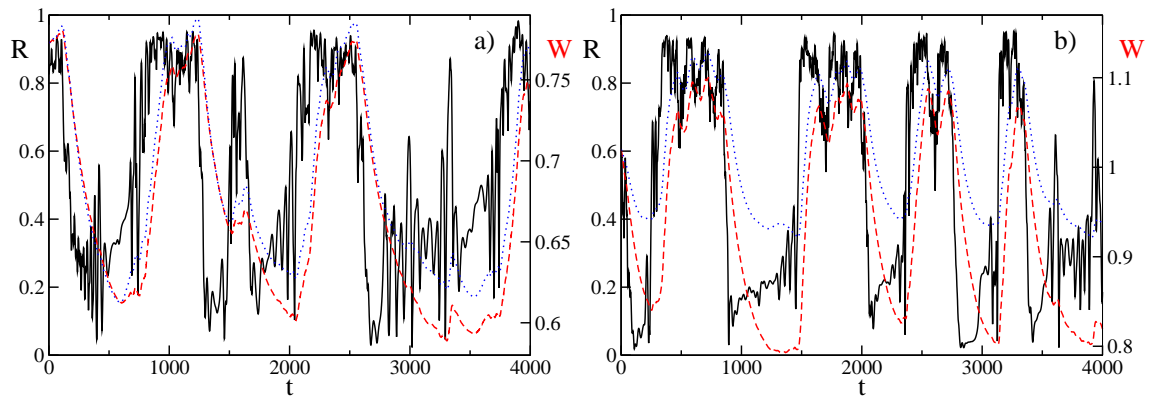

\includegraphics*[angle=0,width=7.5cm]{fig5a} 
\includegraphics*[angle=0,width=7.5cm]{fig5b}
\caption{(color online) Time evolution of $R$
(black solid line) and of $W$ (\textcolor{red}{red} dashed line) for $N = 2000$ neurons. The
dotted (\textcolor{blue}{blue}) line is the mean-field prediction obtained by employing
the map $W(t+\Delta t) = W(t) + \Gamma(t)$, with
$\Gamma$ given by Eq.~\ref{eq:delta}. The data refer  to
$a=1.3, g=0.4$, $d=0.01$; (a) $p=0.01, \alpha=9$: (b) $p=0.02, \alpha=11$
and are measured after a transient $\simeq 10^4$. The mean-field predictions
have been evaluated with $\Delta t=1$, tests performed with $0.5 \le \Delta t \le 40$
did not lead to any peculiar difference.
} 
\label{fig:rw_t}
\end{figure}

\begin{figure}[htb]
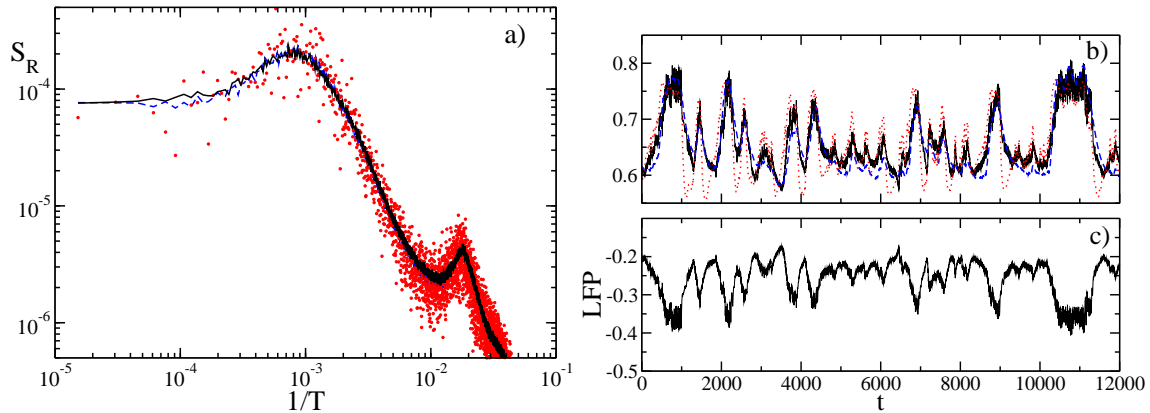

\includegraphics*[angle=0,width=7.5cm]{fig6a}
\includegraphics*[angle=0,width=7.5cm]{fig6b}
\caption{(color online) (a) Power spectrum of the order parameter $S_R$ 
versus the inverse of the period, for $a=1.30 $. The (black) solid line
refers to $N=100$, the (\textcolor{blue}{blue}) dashed line to $N=200$ and the (\textcolor{red}{red}) dots
to $N=400$. The power spectra have been obtained by the time trace of the parameter $R$
sampled at regular intervals $\delta T = 1$ and for a time window
$T_{tot} = 65,536$, the spectra are averaged over 200-400 different time windows 
at $N=100$ and 200 over 10 windows for $N=400$. (b) Time evolution of the filtered 
order parameter $R_f(t)$ (red dotted line), of the average synaptic current $I_f(t)$ 
(black solid line) and of the average synaptic weight $W(t)$ (blue dashed line) for $N=500$.
$R_f(t)$ and $I_f(t)$ have been arbitrarily shifted and rescaled to enhance the similarities
with $W(t)$. (c) LFP as defined in (\ref{LFP}) as a function of time.
The reported results refer to  $g = 0.4$, $\alpha=9$, $d=p=0.01$, 
$\tau_- = 3\tau_+ = 0.3$ and $w_{max}=2$ and all the data 
are obtained after discarding a transient $\ge 10^5$.
} 
\label{fig:spec_filter}
\end{figure}

However, the most interesting dynamical behavior can be observed for intermediate values of the 
intrinsic excitability, namely we will focus on $a=1.3$. As shown in Fig. \ref{fig:rw_t}, 
for this value of the excitability the order parameter $R$ widely fluctuates in 
time from $R \simeq 0$ to $R \simeq 1$, thus indicating that the system jumps irregularly 
between desynchronized and highly synchronized phases. 
This behavior is also observable for the AC as shown in Fig.~\ref{fig:rw_t} (b). The evolution of $R$ 
reveals {\it Low Frequency Fluctuations} (LFFs) 
on time scales of the order of $1,300 \pm 400$ for the SC ($\simeq 700$ for
the AC, data not shown), as it can be clearly appreciated by the corresponding
power spectrum reported in Fig.~\ref{fig:spec_filter} (a). In addition $S_R$ exhibits also a small
subsidiary peak at $T \simeq 50-60$ for the SC ($\simeq 45$ for the AC) indicating that the HFFs are 
still present. It is remarkable that low frequency oscillations
are associated to the relaxation period $\simeq \lambda^{-1}$ previously identified for $a=1.1$.
This seems to suggest that by increasing the neuronal excitability the relaxation process 
becomes an oscillatory one. Somehow the increased excitability is now capable to sustain 
slow collective oscillations, which however for larger $a$ disappears (as shown in Fig. \ref{fig:spectra} (b)). 
Furthermore, the LFFs  for to the AC occurs definitely on a faster time scale.

Let us now estimate the physical time scales over which the 
observed LFFs take place, by assuming a membrane time constant $\tau \simeq 30 - 40$ ms 
(see Appendix B for the conversion units), the frequencies of the slow oscillations are $\simeq 0.02 -0.03$ Hz
therefore they correspond to infraslow rhythms as reported in Ref.~\cite{monto2008}.
On the other hand, the observed HFFs occur in a range of frequencies at the border
between slow and infraslow waves, namely $0.16 - 1$ Hz~\cite{van2011slow}.

To better characterize this regime we report in Fig.~\ref{fig:rw_t} also the time evolution of the 
average synaptic weight $W(t)$, as defined in Eq.~(\ref{ave_w}). This quantity also reveals
low frequency oscillations similar to those of $R(t)$, but occurring with some time delay.
This suggests that $W(t)$ is somehow following the dynamics of $R(t)$, but
the absence of HFFs in the dynamics of $W(t)$ reveals that the average synaptic weight
reveals a sort of {\it inertia}, since it does not respond on short time scales to the modifications 
of the level of synchronization in the network.

To better analyze these similarities let us consider the low-pass filtered
order parameter $R_f(t)$ defined in (\ref{filter}) for the SC. The comparison
with $W(t)$ reported in Fig.~\ref{fig:spec_filter} (b) clearly suggest 
that these two quantities are correlated in time. However, an
almost perfect correlation is observable by considering the filtered synaptic 
current $I_f(t)$ as defined in (\ref{LFP}). In this case $I_f(t)$ and $W(t)$ display  an almost 
identical time evolution (see Fig.~\ref{fig:spec_filter} (b)) thus revealing a strong
correlation among the synaptic weights and the synaptic currents, at least
by considering the corresponding (ensemble) averaged quantities.
 
In Fig.~\ref{fig:spec_filter} (c) we reported also the corresponding LFP 
(see (\ref{LFP})). This exhibits minima with superimposed high frequency oscillations 
in the high activity phase, corresponding to the high synchronization, and maxima
in the low synchronized phase. Despite this trace resembles strongly 
the spontaneous fluctuations observed in cortical activity of mammals during 
slow-wave sleep or during quiet wakefulness~\cite{harris,chauvette2011}, 
we should remark that our {\it up-states} and {\it down-states} are characterized 
by a tonic firing of neurons with similar average ISIs.
\footnote{We observe a variation of less than 10\% of the ISI in the two states.}
In our case, the difference among these two states is mainly
in the degree of synchronization of the population activity which is high (low)
in the up-phase (down-phase). At variance with the experimentally observed activity in 
synchronized cortical states, typical of slow-wave sleep and quite wakefulness,
which is characterized by up-phases (down-phases) associated to
high level (absence) of firing activity~\cite{harris}.

\section{Effective free energy landscape}

\begin{figure}[htb]
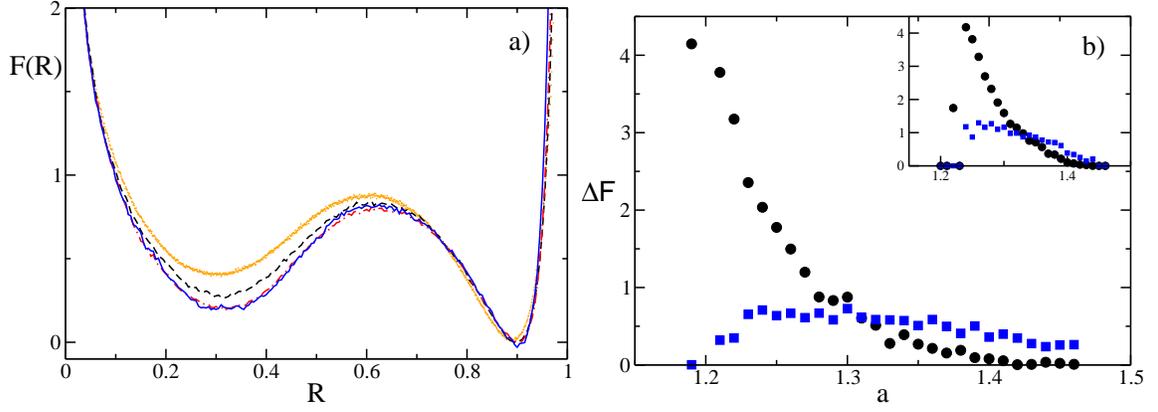

\includegraphics*[angle=0,width=7.5cm]{fig7a}
\includegraphics*[angle=0,width=7.5cm]{fig7b}
\caption{(color online) (a) Free energy landscape $F(R)= - \log P(R)$
as obtained from the PDF of the order parameter $P(R)$ for $N=50$
(\textcolor{orange}{orange}) dots,
$N=100$ (black) dashed line, $N=200$ (\textcolor{red}{red}) dot-dashed line and $N=500$ (\textcolor{blue}{blue}) solid line.
The data are obtained by considering
trajectories of duration $\simeq 2.7\times10^6 - 5.3 \times 10^6$ sampled
at regular interval $\delta T =1$, after discarding a transient $\simeq 10^7$ spikes.
(b) Free energy barriers $\delta F$ as a function of the intrinsic excitability $a$:
the $\CIRCLE$ (\textcolor{blue}{$\blacksquare$}) denote the barrier $\Delta F$ 
separating the minimum $R_H$ ($R_L$) from the maximum $R_S$ for a network of $N=500$ neurons.
The reported results refer to  $g = 0.4$,  
$\tau_- = 3 \tau_+ = 0.3$, $w_{max}=2$, $\alpha=9$ and $d=p=0.01$, apart in the inset of panel (b)
where we considered $p=2d=0.02$ and $\alpha=11$.
} 
\label{fig:pot}
\end{figure}

A further characterization of this regime can be achieved by considering the probability density 
distribution $P(R)$ of the order parameter obtained by measuring $R$ at regular time intervals
$\delta T$ during a long simulation, after discarding an initial transient time.
In particular, we prefer to visualize the obtained results in term of the corresponding 
effective {\it free energy} landscape, as defined by $F(R) = -\log P(R)$\footnote{Here and in the following we assume
an unitary scale for the energy.}, as plotted  in Fig.~\ref{fig:pot} (a). 
$F(R)$ reveals two minima at $R_L \simeq 0.3$ and $R_H \simeq 0.9$
corresponding to the LS and the HS regime, respectively. These two states are separated by
a maximum located at $R_S \simeq 0.6$. The HS minimum is more pronounced and separated by a higher
barrier $\Delta F$ from the saddle at $R_S$, this indicates that the system spends
more time in the HS regime and that this state is characterized by a quite well defined level of 
synchronization. On the other hand the LS state corresponds to a broader minimum, reflecting
the fact that the system in the LS regime visits states with levels of synchronization distributed
over a broader range than in the HS state. From Fig.~\ref{fig:pot} (a) it is also evident
that the large oscillations between LS and HS measured by $R(t)$ do not vanish in 
the thermodynamic limit, since  $F(R)$ tends to an asymptotic profile already for $N > 100$.

The level of stability of the HS (LS) state can be measured in terms of the free energy
barrier $\Delta F_{H}$ ($\Delta F_L$) separating $R_H$ ($R_L$) from the saddle $R_S$. These data are
reported in Fig.~\ref{fig:pot} (b) for $d=p=0.01$ and $\alpha=9$ 
(and for $p=2d=0.02$ and $\alpha=11$ in the inset) for a wide range
of intrinsic excitability. We observe for the symmetric (asymmetric) case
that finite barriers for both states exist only
for $a \in [1.19;1.46]$ ($a \in [1.22;1.45]$), therefore in this interval HS and LS regime coexist.
For $a \to 1.18$ ($a \to 1.21$) the HS barrier appears to diverge thus indicating that for smaller
$a$-values the system is fully synchronized, while for $a \ge 1.48$ ($a \ge 1.47$)
the two minima
merge (the associated barriers vanish) in an unique LS state. 
These results are consistent with the analysis reported in Fig.~\ref{fig:r_a} (a).

\begin{figure}[htb]
\includegraphics*[angle=0,width=7.5cm]{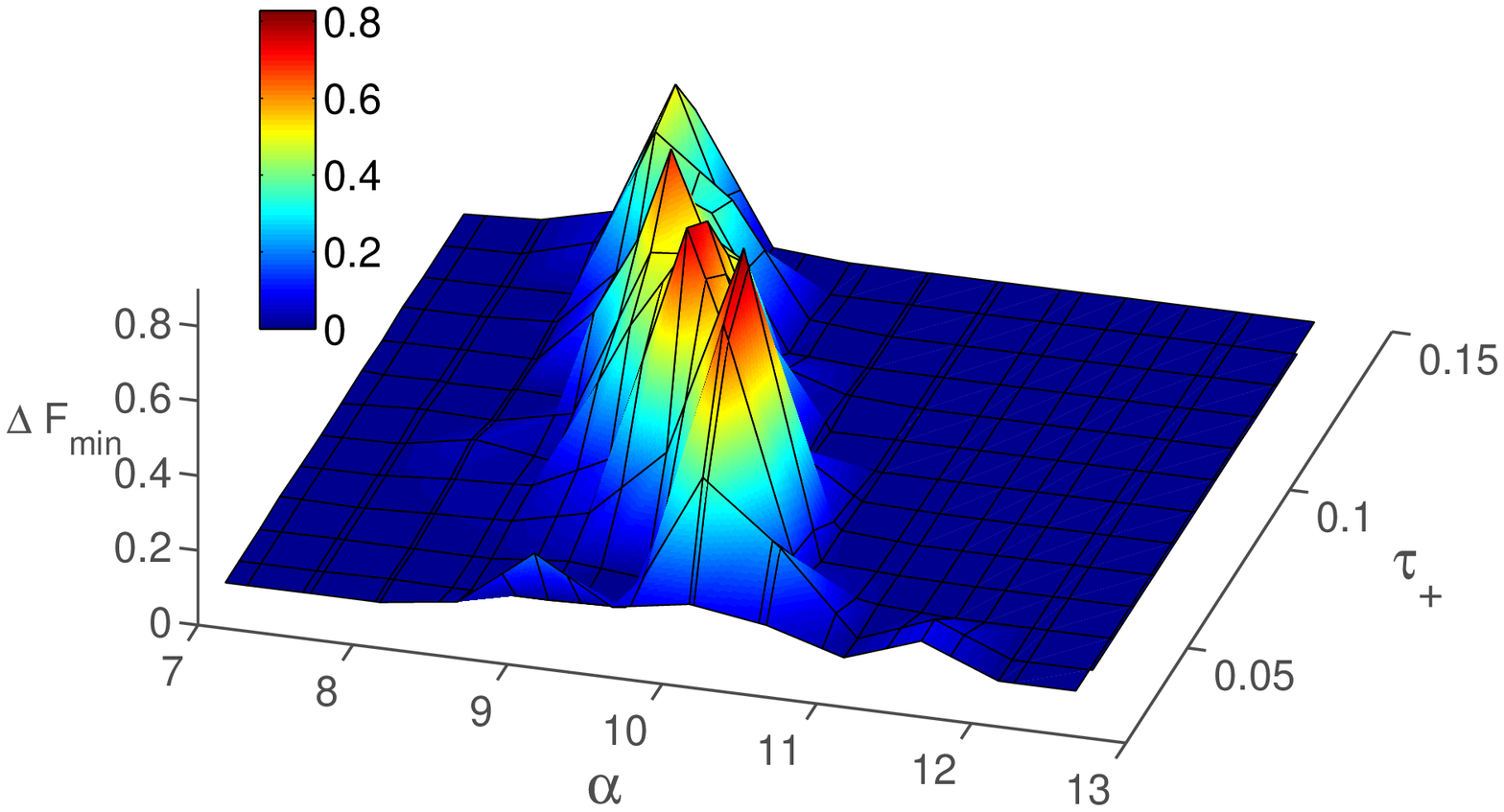}
\includegraphics*[angle=0,width=7.5cm]{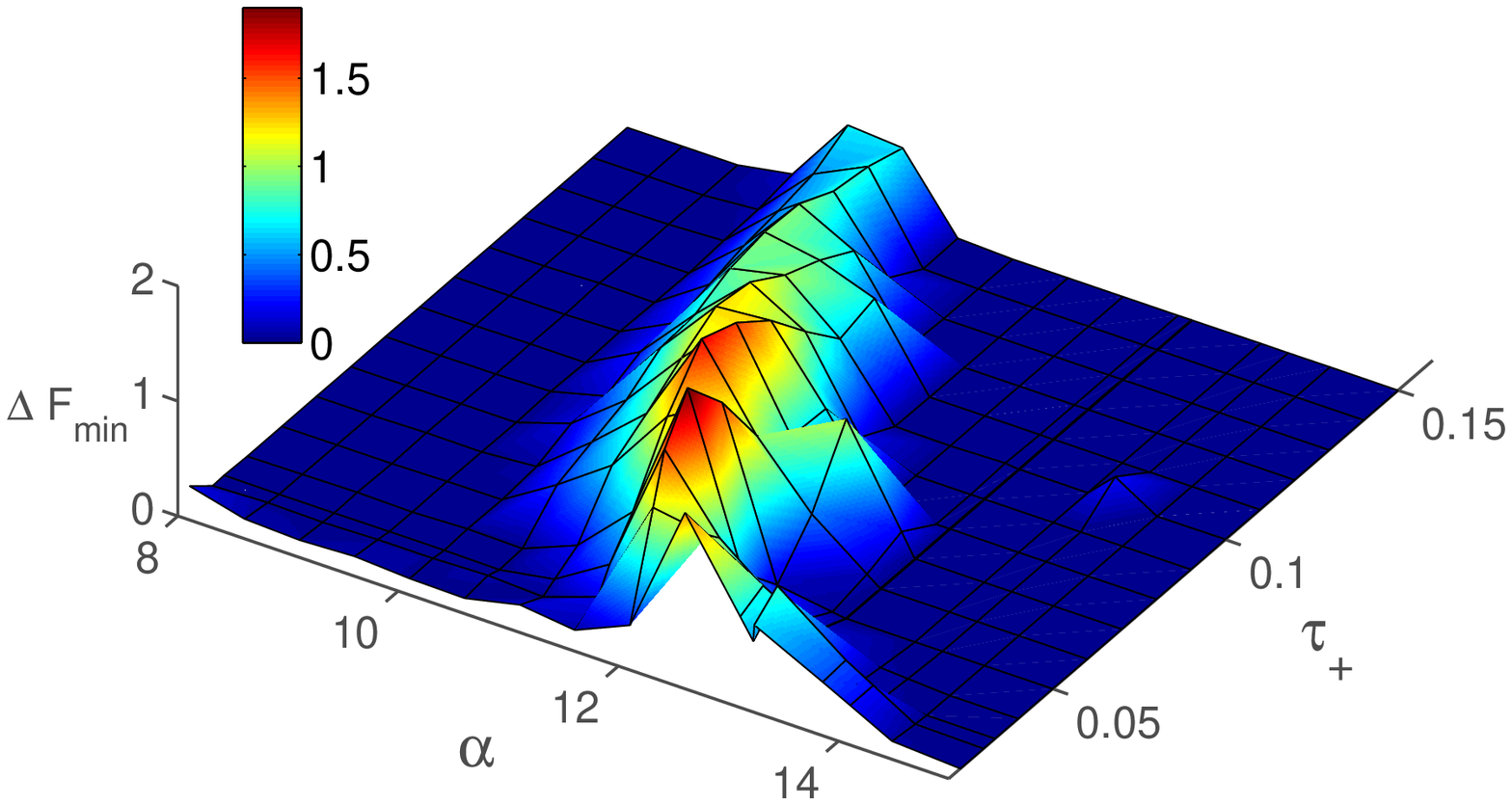}
\caption{(color online) Minimal free energy barriers $\Delta F_{min}$ as a function of 
$\alpha$ and $\tau_{+}$ for $p=d=0.01$ (a) and $p=2  d=0.02$.
The barrier heights are color coded, the scale is reported next to the corresponding figure.
The reported results refer to  $g = 0.4$, $a=1.3$, $N=200$, $\tau_- = 3 \tau_+ $, and $w_{max}=2$.
} 
\label{fig:barriers}
\end{figure}

To determine if the observed oscillations between HS and LS persist by varying the
relevant time scales (i.e. the synaptic time scale $\alpha^{-1}$ and the learning
time windows),  we have analyzed a large interval in the $(\alpha,\tau_+)$-plane.
In particular, as suggested by the experimental evidences we fixed 
$\tau_- = 3 \tau_+$ and we examined both the cases $d=p=0.01$ and $p=2 d = 0.02$,
with the other parameters held constant to $g=0.4$ and $a=1.3$.
For each parameter set we have estimated the minimal free energy barrier
$\Delta F_{min}$ between  $\Delta F_{H}$ and $\Delta F_L$, whenever $\Delta F_{min}$
is non zero this means that the neuronal dynamics oscillates between high synchronous
and a low synchronous states, and therefore LFF oscillations are present.
As shown in Fig.~\ref{fig:barriers}, the barriers are finite only in a limited stripe of
the $(\alpha,\tau_+)$-plane, for smaller (larger) $\alpha$-values the system is in the
LS (HS) regime. In particular, for $0.02 \le \tau_+ \le 0.15$ one has $\Delta F_{min} >0$
within the interval $\alpha \in [8;10]$  ($\alpha \in [10;12]$) for  $d=p=0.01$
($p=2 d = 0.02$). Thus suggesting that the observation of LFFs is limited to
synaptic rise/decay time $1/\alpha \simeq 0.1$ and it depends only slightly
on $\tau_+$, at least in the examined range: by increasing $\tau_+$ by a factor 
5 the corresponding synaptic time needed to observe a finite  $\Delta F_{min}$ 
grows only by 20\%.

\section{Constrained Simulations}

\begin{figure}
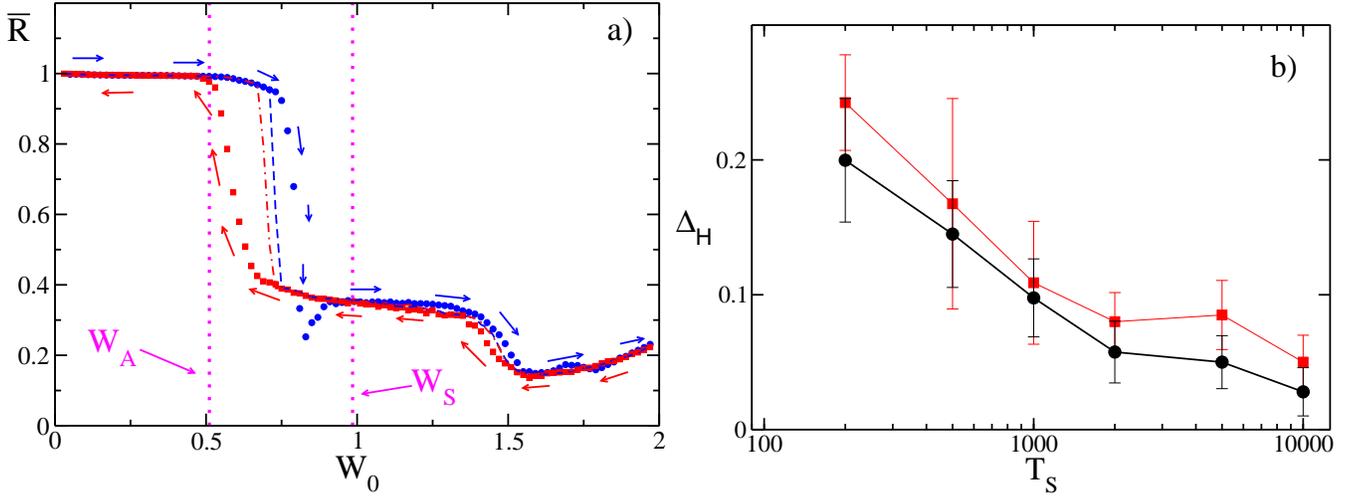

\includegraphics[width=0.49\textwidth]{fig9a}
\includegraphics[width=0.49\textwidth]{fig9b}
\caption{(Color online) (a) Plots of $\bar R$ vs $W_0$ as obtained during constrained simulations of duration $T_S$
for increasing (\textcolor{blue}{$\CIRCLE$} and \textcolor{blue}{-~-}) and decreasing (\textcolor{red}{$\blacksquare$} and \textcolor{red}{$\cdot -$}) $W_0$. 
The symbols correspond to $T_S=200$ and the lines to $T_S=10,000$. The magenta dotted vertical lines 
indicate the fixed point values $W_A$ and $W_S$. Results averaged over $100$ different initial conditions
and $\delta W_0 = 0.02$. The data refer to $p=0.01$ and $\alpha=9$. 
(b) Width of the hysteretic loop $\Delta_H$ as a function of $T_S$ for $p=0.01$, $\alpha=9$ 
(black circle) and $p=0.02$, $\alpha =11$ (red squares), the error bars have been estimated 
by measuring $\Delta_H$ for $250$ ($33$) different initial conditions, respectively.
All the data are for $d=0.01,\tau_-=3\tau_+=0.3, g=0.4, a=1.3, N=200$.}
\label{fig:lockedTransition}
\end{figure}

\noindent 
As previously noticed it seems that, at least at a mean field level, 
the dynamics of the synaptic weights and the synchronization of the firing events
are somehow related. In order to clarify this relationship, if any, we 
will examine the neuronal population dynamics {\it decoupled} from the evolution of the average 
synaptic strength by performing CSs. During this
kind of simulation the average synaptic weight is maintained constant and equal to a fixed value $W_0$.
In particular, we set initially $W_0=0$ and follow the evolution of the system for a time span $T_S$.
Then, we perform a new simulation of duration $T_S$, starting from the last configuration of 
the previous run, with an increased synaptic weight $W_0=\Delta W_0$. We repeat
this procedure by increasing $W_0$ at regular steps $\Delta W_0$ until 
$W_0$ reaches the maximal allowed value, namely $w_{M}$. 
By applying the reverse protocol, $W_0$ is successively decreased (always at steps of $\Delta W_0$) 
until it returns to zero. During each single simulation we measure the average order parameter $\bar R$
only over the second half of the run (therefore on a time interval $T_S/2$),
this in order to allow the system to relax after each modification of $W_0$. 
The corresponding results are shown in Fig.~\ref{fig:lockedTransition} (a) for $T_S=200$ and $T_S=10,000$.  In this manner, $W$ is held fixed, while the individual $w_{ij}$ are essentially free to evolve.

At low $W_0$ the system is fully synchronized $\bar R \simeq 1$,
by increasing the average synaptic weight the order parameter drops to a LS state above a 
critical value $W_0^{(1)}$. Furthermore,the system resynchronizes, by decreasing $W_0$, at a lower
synaptic weight value, namely $W_0^{(2)} < W_0^{(1)}$. This seems to indicate that the transition
is discontinuous and hysteretic, however by increasing from $T_S=200$ to $T_S=10,000$ the transition 
remains discontinuous, but the width of the hysteretic loop 
$\Delta_H = W_0^{(2)} - W_0^{(1)}$ shrinks noticeably, as evident from Fig.~\ref{fig:lockedTransition} (a).
To better investigate this point, we report $\Delta_H$ as a function of $T_S$ in 
Fig.~\ref{fig:lockedTransition} (b), these data seem to suggest that $\Delta_H$
will vanish for adiabatic transformations of  $W_0$, corresponding to $T_S \to \infty$.

However, since we are interested in
characterizing the transition between HS and LS observed during LFFs
and since each oscillation in $R(t)$ ($W(t)$) takes place on a finite time
the limit $T_S \to \infty$ is not of interest for this analysis.
However, it is not trivial to estimate which is a meaningful $T_S$-value 
to employ in the CSs to compare the obtained results with those of the 
corresponding USs. A first constraint on $T_S$ is that it should be 
sufficiently long with respect to the period of the HFF, this in order to 
get rid of the fast oscillations of $R(t)$ during CSs. On the other hand,
the variation of $W_0$, performed during CSs, should be done on time scales
of the order of the period of the LFFs.  
To be more specific, let us focus on the SC. For these parameter values, 
HFFs occur on time scales $T_{HFF} \simeq 50-60$, therefore the first request 
is that $T_S >> T_{HFF}$.  Furthermore, the period of the LFFs is 
$T_{LFF} \simeq 1,000 - 2,000$. For the AC, the time scales 
associated to LFFs and HFFs are faster.
As a matter of fact, to be on the safe side 
we have employed $T_S \simeq 200 - 1,000$.

\section{Mean-field prediction of synaptic weight dynamics }

\noindent 
We now address the influence of the level of synchrony of the neuronal population,
as measured by the order parameter $R$, on the synaptic weight dynamics, using a mean-field 
analysis for $W$. In particular, we examine the dynamics of $W$
in the two extreme cases of fully synchronized and asynchronous evolution of the network.

From Eqs.~(\ref{wrule},\ref{post}) and by following the approach described in ~\cite{Izhikevich2003BCM},
one can obtain the average synaptic weight modification $\Gamma$, associated to
each presynaptic spike, 
\begin{equation}
\Gamma(t)= p(w_{M}- W) \int_0^\infty d\delta P(\delta) 
{\rm e}^{\frac{-\delta}{\tau_{+}}} 
- d  W   \int_{-\infty}^0 d\delta P(-\delta) 
{\rm e}^{\frac{\delta}{\tau_{-}}}
\label{eq:delta}
\end{equation}
where $P(\delta)$ is the PDF of the time differences $\delta$ 
between postsynaptic and presynaptic firing times. 
To test the predictive value of Eq.~(\ref{eq:delta}), we have measured from USs
$P(\delta)$ at regular time intervals $\Delta t$, thanks to these data 
we can estimate the evolution of the synaptic weight at regular time intervals, 
as $W(t+\Delta t) = W(t) + \Gamma(t)$. This reconstruction gives a quite good
estimation of the true evolution for SC (see the dotted
blue line in Fig.~\ref{fig:rw_t} (a)). However the agreement
declines for the AC, in this case the mean-field evolution
still catches the oscillations of $W(t)$ with the correct periods, 
but it overestimates the minimal values reached by
the synaptic weights, as shown in Fig.~\ref{fig:rw_t} (b).
We have verified that the mean field prediction remains good
in the symmetric case $p=d$, even by doubling
the value of $p$, and also by considering a situation with depression
prevailing on potentiation, namely $d= 2 p$. 
The origin of the partial failure of the mean-field prediction (\ref{eq:delta})
could be related to the fact the collective dynamics becomes faster in the AC, 
as discussed in Sect. III.  Despite this, 
the overall picture seems still to work also when potentiation is larger than depression, 
and the mechanism at work for the generation of LFF
seems the same in the SC and AC, as detailed in Sect. VII.

\begin{figure}
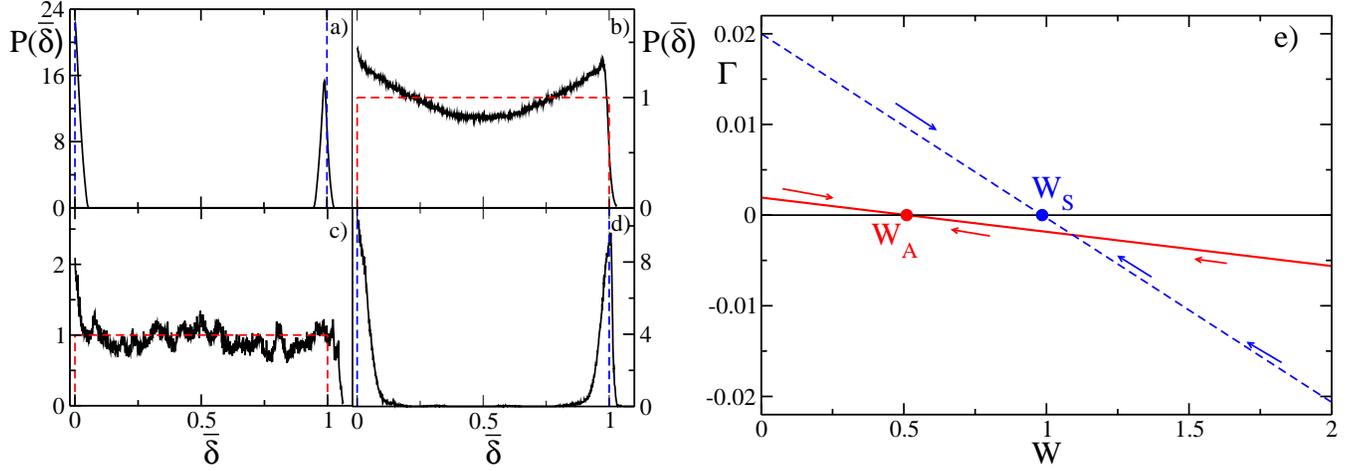

\includegraphics[width=.51\textwidth]{fig10a}
\includegraphics[width=.47\textwidth]{fig10b}
\caption{(color online) 
Probability distribution functions $P(\bar \delta)$ as as a function of the
renormalized time differences $\bar \delta_{ij} = \delta_{ij} / <ISI>$,
where $<ISI>$ is the average inter-spike interval measured for the considered
state. The reported PDF refer to $a=1.09$ (a);  $a=1.7$ (b) and $a=1.3$ (c) and (d).
For the latter case, the distributions have been obtained during an unconstrained simulation 
by considering the $N \times (N-1)$ $\delta_{ij}$ values associated to the last $N$ spikes
preceding a strongly (weakly) synchronized state corresponding to
an order parameter value $R \simeq 0.11$ (c) ($R \simeq 0.98$ (d).
The \textcolor{blue}{-~-} refer to $P_S$ in (a) and (d), while
the \textcolor{red}{-~-}  refer to $P_A$ in (b) and (d).
(e) $\Gamma$ versus $W$ for the asynchronous and the synchronized
case: namely $\Gamma_A$ (\textcolor{red}{--}) and $\Gamma_S$ 
(\textcolor{blue}{-~-}). The arrows denote the direction of the evolution of
$W(t^+)$ due to the modifications induced by $\Gamma(W(t^-))$. 
The \textcolor{red}{$\CIRCLE$} (\textcolor{blue}{$\CIRCLE$}) indicates $W_A \simeq 0.51$ ($W_S \simeq 0.985$).
The other parameters of the simulations are $\alpha=9$, $\tau_- = 3 \tau_+ = 0.30$, $d=p=0.01$,
$g=0.4$ and the network size is $N=500$}
  \label{fig:Pdelta}
\end{figure}

By assuming that the postsynaptic neuron is firing with period $T_0$, we are able to derive the time difference distribution $P(\delta)$ for the two limiting cases: fully synchronized and asynchronous dynamics. In the fully synchronized (asynchronous) situations we expect a distribution of the form $P_S(\delta)= {\cal D}(\delta) +
{\cal D}(\delta-T_0)$ ($P_A(\delta)= 1/T_0$) defined in the interval $[0, T_0]$. Here ${\cal D}$ denotes a Dirac delta function. These guesses are essentially confirmed by direct USs as shown in Fig. \ref{fig:Pdelta}. In particular,
the data reported in Fig. \ref{fig:Pdelta} (a) (Fig. \ref{fig:Pdelta} (b)) refer to a high (low) synchronized
state corresponding to $a=1.09$ ($a=1.7$). On the other hand the results shown in Fig. \ref{fig:Pdelta} (c) and
(d) refer to the same state, corresponding $a=1.3$, where $P(\delta)$ is measured 
by considering the $\delta_{ij}$ values associated to the last $N$ spikes
preceding a strongly (weakly) synchronized phase with associated an order parameter value 
$R \simeq 0.11$ (c) ($R \simeq 0.98$ (d)). Therefore, at least in these two cases,
we can derive an analytic estimation of $\Gamma$. By assuming that the post-synaptic neuron 
fires with constant period $T_0$ we can perform the integrals appearing in (\ref{eq:delta})
obtaining the following results.

\subsection{Asynchronous dynamics}

\noindent In this situation $P(\delta) = P_A (\delta)$ and we can rewrite (\ref{eq:delta}) as follows
\begin{equation}
\Gamma_A =  \frac{1}{T_0} \left[ p\tau_+(2-\w)\left(1-e^{-T_0/\tau_+}\right) -d\w\tau_-\left(1-e^{-T_0/\tau_-} \right)\right]  \quad.
\end{equation}
As shown in Fig. \ref{fig:Pdelta} (e), $\Gamma_A$ vanishes for $W=W_A$ and it is positive
(negative) for $W < W_A$ ($W > W_A$), therefore for the dynamics of $W(t)$
\be 
\w_A = \frac{2P}{S+P} \ , \ P = \tau_+\left(1-e^{-T_0/\tau_+}\right) \ , \ D = \tau_-\left(1-e^{-T_0/\tau_-}\right) \label{eq:WA} 
\ee
is a stable fixed point. The value of $W_A$ only depends  on the STDP parameters and $T_0$,
by assuming that the period $T_0$ is equal to the average inter-spike interval, we can estimate the value of this fixed point. As shown
in Fig.~\ref{fig:lsweep_p1p2} this prediction gives a good estimation of the $W$ values
in the low synchronized regime (corresponding to $ a > 1.6$) both for $p=d$ as well as for $p = 2 d$.

\subsection{Fully synchronized dynamics}

\noindent For the fully synchronized situation $P(\delta) = P_S (\delta)$ and (\ref{eq:delta}) becomes

 \begin{equation}
\Gamma_S = p(2-\w)\left(1+e^{-T_0/\tau_+}\right)-d\w\left(1+e^{-T_0/\tau_-}\right)
\quad .
\end{equation}
As reported in Fig. \ref{fig:Pdelta} (e), $\Gamma_S$ vanishes 
for $W=W_S$ and it is positive (negative) for $W < W_S$ ($ W > W_S$), therefore the solution 
\be 
\w_S = \frac{2\tilde{P}}{\tilde{P}+\tilde{S}} \ , \ \tilde{P} = 1+e^{-T_0/\tau_+} \ , \ \tilde{D} = 1+e^{-T_0/\tau_-}
\quad ,
 \label{eq:WS} 
\ee
represents an stable fixed point. Also in this situation by setting $T_0 = < ISI>$ we can
obtain a numerical estimation of $\w_S$, as clearly shown in Fig.~\ref{fig:lsweep_p1p2}. This
represents an upper bound for $W$ for any studied regime, while $W  \to \w_S$ only 
for $a \to 1$, corresponding to the state of maximal synchronization.

 To summarize, in both cases $\Gamma$ vanishes for a finite value of the average synaptic weight, 
namely $W_S$ ($W_A$) for the synchronized (asynchronous) situation. Additionally, for $W < W_S$ ($W > W_A$) 
the synapses are on average potentiated (depressed), while a similar mechanism rules for the 
asynchronous case. This implies that  $W_S$ ($W_A$) is a stable attractive point for 
the dynamics of $W$ in the synchronized (asynchronous) regime. All these results
apply in the symmetric (asymmetric) case $p=d = 0.01$ ($p=2d = 0.02$), whenever $\tau_- = 3 \tau_+ = 0.30$.
However, in general the system will not be completely synchronized
or asynchronous and the distribution $P(\delta)$ will lie in between the 
extreme cases represented by $P_S$ and $P_A$, therefore we expect that 
the values of $W$  will be bounded within the interval $[W_A, W_S]$.
This expectation is fully confirmed for $p=d = 0.01$, as shown in Fig.~\ref{fig:lsweep_p1p2} (a),
while for $p=2d = 0.02$ $W < W_S$ for any value of the intrinsic excitability $a$, but in the intermediate
regime (namely, $ 1.25 <  a < 1.50$) $W$ can attain values somehow lower than $W_A$.
This result is due to the not perfectly good predictive power of the mean field approach
in this specific case, as already stated previously.

\section{Sisyphus Effect}

We have now all the ingredients needed to explain the LFFs reported previously and to
single out the mechanism responsible for such behavior.
Let us suppose that initially the system is in the HS phase 
with an associated average low coupling value $W < W_0^{(1)}$. 
In this regime the attractive fixed point $W_S$ is larger
than the the transition point $W_0^{(1)}$ (see Fig. \ref{fig:lockedTransition} (a)). 
Therefore $W$ increases and tends towards $W_S$, until $W > W_0^{(1)}$,
at which point the system starts to desynchronize and to approach the LS state,
the value of $R$ dropping. In this desynchronizing stage
the distribution $P(\delta)$ becomes almost flat (see  Fig. \ref{fig:Pdelta} (c))
and the attractive point for the synaptic evolution will now be $W_A$, 
which is located below $W_0^{(2)}$ (as shown in Fig. \ref{fig:lockedTransition} (a)).
The synaptic plasticity decreases $W$ in order to reach $W_A$, but when the average synaptic 
weight crosses $W_0^{(2)}$ the neurons begins to resynchronize. This brings
the system back to the HS state from where it started and the cycle
can now be repeated. The cycle will continue indefinitely, and is the essence of the 
Sisyphus Effect.

One should remark that the above arguments are approximate, because the system 
is never exactly fully synchronized or desynchronized. Instead the network passes 
through a continuum of states, each associated with a different fixed point in $W$-space. 
The crucial ingredient for the emergence of the SE 
is that the fixed points corresponding to the HS (LS) phase
are larger than the transition point $W_0^{(1)}$ (smaller than  $W_0^{(2)}$). 
As we have verified that this is indeed the case, the described mechanism can 
be considered as still effective.

\begin{figure}[htb]
\includegraphics[width=.5\textwidth]{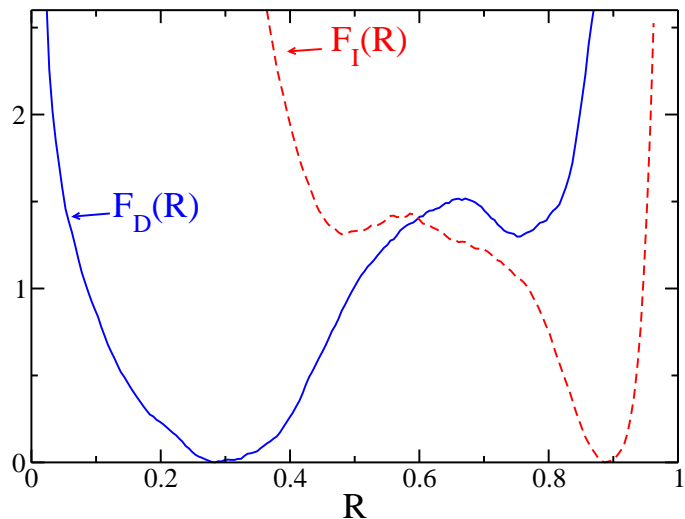}
\caption{(color online) Conditional free energy profiles for $R$: \textcolor{blue}{--} refers
to $F_D(R)$, while \textcolor{red}{-~-} to $F_I(R)$. The PDF has been obtained during USs
by averaging over measurements of $R$ obtained at regular time intervals $\Delta T =1$, 
The data correspond to $N=2000,\alpha=9,g=0.4,a=1.3,p=d=0.01,\tau_-=3\tau_+=0.3$.}
\label{fig:conditional_energy}
\end{figure}

To perform a further test of the validity of our analysis, we measure the probability distribution function (PDF)
of the order parameter $R$ conditioned to the fact that $W$ was increasing (decreasing) during an US.
From the PDFs we can derive  the corresponding free energy profiles $F_I(R)$ and $F_D(R)$, which are
reported in Fig. \ref{fig:conditional_energy}. From the figure $F_I$ has a principal minimum at $R_H$,
while $F_D$ has an absolute minimum at $R_L$. Both profiles reveal a shoulder at intermediate $R$ values.
These results confirm that the equilibrium attractive values for $W$ are located opposite to the 
transition points, because when the system is in the HS (LS) regime the synaptic weights increase 
(decrease) continuously trying to reach the corresponding fixed points.

\begin{figure}[htb]
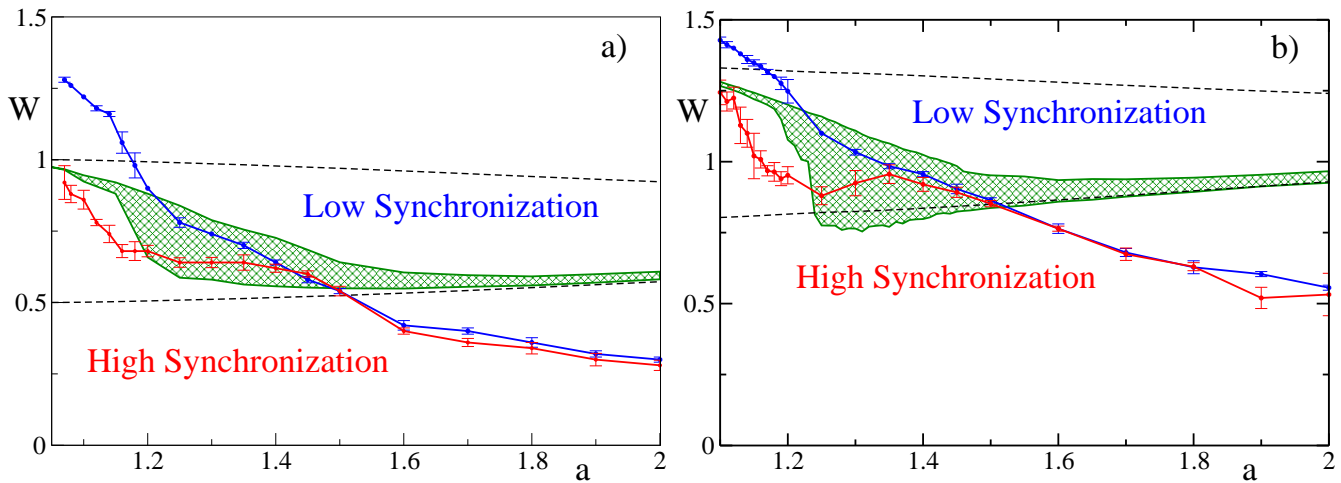

\includegraphics[width=0.49\textwidth]{fig12a}
\includegraphics[width=0.49\textwidth]{fig12b}
\caption{(color online) Average synaptic plasticity $W$ versus intrinsic excitability $a$.
The \textcolor{green}{shaded area} indicates the distribution of the $W$ values measured during USs.
The upper \textcolor{blue}{blue} (low \textcolor{red}{red}) line represents $W_0^{(1)}$ ($W_0^{(2)}$)
The error bars have been evaluated over 5 different realizations
of the CSs. The upper (lower) dashed black line represents the fixed point values $W_S$ ($W_A$).
The data refer to (a) $p = d$ and $\alpha=9$ ; (b) $p=2d$ and $\alpha=11$; while the other parameters are set
to $N=200$, $g=0.4$, $T=1000$, $d=0.01$,$\tau_-=3\tau_+=0.3$.}
\label{fig:lsweep_p1p2}
\end{figure}

As a final analysis, we have verified the validity of the SE over a wide
range of intrinsic excitabilities for the symmetric and asymmetric case.
We expect that the SE appears whenever the transition values
$W_0^{(1)}$ and  $W_0^{(2)}$ lye within the interval $[W_A, W_S]$.
Therefore, we have measured the transition values $W_0^{(1)}$ and  $W_0^{(2)}$
and the corresponding fixed points for a large interval of intrinsic excitabilities,
namely $1 < a \le 2$ for $p = d = 0.01$ and $p=2d= 0.02$ (see Fig.~\ref{fig:lsweep_p1p2}).
For, $p=d$ ($p=2d$) the transition is hysteretic in the interval $a \in ]1, 1.40]$
($a \in ]1,1.45]$) while for larger $a$-values $W_0^{(1)}$ and  $W_0^{(2)}$ are essentially coincident.
Furthermore, for $a \le 1.18$ ($a \le 1.17$)  $W_0^{(1)} \ge W_S$, while
for $a \ge 1.50$ ($a \ge 1.51$) $W_A \ge W_0^{(1)}, W_0^{(2)}$.
Furthermore, the $W$ distributions measured during USs are reported in
Fig~\ref{fig:lsweep_p1p2} as shaded areas and they include the transition
interval $[W_0^{(1)},W_0^{(2)}]$ for $ 1.20 \le a \le 1.48$
($ 1.20 \le a \le 1.51$). In the AC one observes that, in the range of parameters where the
Sisyphus mechanism is at work, the measured
$W$ can be smaller than $W_A$, instead outside this region 
$W_A$ always represents a lower bound for the distribution of the $W$.
This is in line with what previously reported in Sec. VI concerning the
predictive value of Eq.(\ref{eq:delta}) and it points out that 
one should go beyond the mean-field approximation to get 
a better reproduction of the $W$ dynamics, at least in the AC.

As previously shown in Fig.~\ref{fig:pot} (b) the free energy $F(R)$ reveals
the coexistence of two minima, corresponding to the competing HS and LS states,
within the interval $a \in [1.19:1.46]$ ($a \in [1.22:1.45]$) for the symmetric
(asymmetric) case. These results clearly indicate that the Sisyphus Effect
is responsible for the LFFs observed in the dynamics of our network.

\section{Conclusions}

In this paper we have reported a simple deterministic mechanism, the Sisyphus Effect,
responsible for the onset of irregular collective oscillations between asynchronous
and synchronous states in excitatory neural networks with STDP. 
The transition between the two states are driven by STDP: the synaptic
weights tend to relax towards their equilibrium values, which in turn are
determined by the synchrony in the neural population.

For intermediate values of the synaptic characteristic time,
the system is fully synchronized for sufficiently small synaptic weights,
while it becomes asynchronous above a critical coupling.
However, for synchronized (desynchronized) neural activities
the synaptic weights tend towards large (small) equilibrium
values corresponding to asynchronous (synchronous) evolution.
The activity of the network can be schematized as that
of a one dimensional order parameter evolving on a free energy 
landscape displaying two coexisting equilibria. 
Depending on the small (large) values of the synaptic weights 
the landscape is tilted towards the strongly (weakly) synchronized state,
thus becoming the attractive equilibrium for the dynamics.
On the other hand, the synchronized (desynchronized) neural
activity increases (reduces) the weights until their values
force the landscape to tilt in the opposite direction.
This drives both the observables into a never ending cyclic 
behavior. 
 
The SE should be observable in pulse coupled neural networks for usual STDP 
whenever excitation has a desynchronizing effect. On one hand, this is, in general, verified for
any kind of neuronal response (type I or type II) for sufficiently
slow synaptic interactions \cite{Vreeswijk1994,hansel1995}. 
On the other hand, for fast excitatory synapses, we expect
that for temporally inverted STDP rules~\cite{letzkus2006} the Sisyphus effect
should be active. Furthermore, we have shown that this effect is 
present also for biologically relevant choices of the STDP parameters
and the collective oscillations occurs on timescales corresponding to
infraslow oscillations observed in the brain dynamics.
In relation to this, it is very interesting to note that 
the authors of refs. ~\cite{monto2008,vanhatalo2004infraslow} described 
infraslow fluctuations in the excitability of real neural networks. Such fluctuations 
could be explained by the slow oscillations of the synaptic couplings as caused by 
the Sisyphus Effect.

Ultraslow rhythms have been previously reproduced in 
an excitatory network composed of fully coupled conductance based 
neurons~\cite{hlinka2010}. In particular, the authors proposed a mechanism, 
based on retrograde endocannabinoid signaling, which was quite
similar to the Sisyphus Effect. Also in Ref.~\cite{hlinka2010} 
high level of synchrony induced a feedback mechanism
based on the evolution of a mean-field variable which leads to a decoupling of 
the neurons. Furthermore, whenever the neurons desynchronize the suppression mechanism 
was removed and the populations can evolve towards its synchronous activity.
However, at variance with the SE the evolution time scale of the mean
field control parameter was phenomenologically set to be $\simeq 10 - 100$ s.

The reported deterministic collective dynamics is absolutely not trivial, and deserves further
analysis to understand if the observed collective
motion, which is clearly chaotic from a microscopic point of view, can be considered a 
further example of collective chaos, similar to the one recently reported for two 
coupled sub-populations of neurons~\cite{Olmi2010Chaos}.

\begin{acknowledgments}
We acknowledge illuminating discussions with B. Lindner, L. Shimansky-Geier,
M. Wolfrum, A. Politi, Y. Maistrenko, S. Lepri, S. Luccioli, M. Bazhenov, 
J. Berke, S. Coombes, D. Angulo-Garcia.
AT thanks the VELUX Visiting Professor Programme 2011/12
and the Aarhus Universitets Forskningsfond for the support received during his stays
at the University of Aarhus (Denmark).
This work is part of the activity of the Marie Curie Initial  Training Network 'NETT'
project \# 289146 financed by the European Commission and it has
been partially supported by the DFG - Deutsche Forschungsgemeinschaft
in the framework of the Collaborative Research Center SFB 910.
\end{acknowledgments}

\FloatBarrier

\appendix

\section{Synaptic weights distributions}

\begin{figure}[htb]
\includegraphics[width=.5\textwidth]{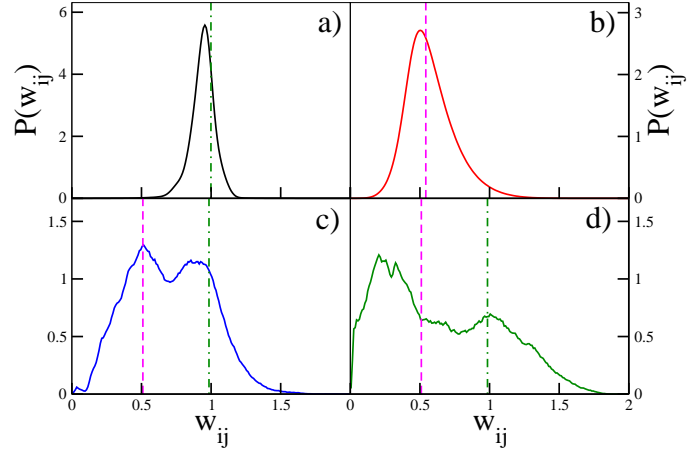}
\caption{(color online) Probability density distributions $P(w_{ij})$ for
$a=1.09$ (a), $a=1.7$ (b) and for $a=1.3$ measured for small (large) synchronization
values $R \simeq 0.1$ ($R \simeq 0.9$) corresponding to panels (c) and (d), respectively.
These latter distributions have been obtained by averaging over $20$ different configurations 
of similar $R$-value (the 20 highest and lowest during a long run, respectively),
while the other two are single snapshot. 
The vertical magenta dashed (green dot-dashed) lines indicate the mean-field
equilibrium points $W_A$ ($W_S$) in the considered cases.
The parameters are $N=500,\alpha=9,g=0.4,a=1.3,p=d=0.01,\tau_-=3\tau_+=0.3$.}
\label{fig:Pw}
\end{figure}

\noindent In this appendix we investigate the shape and the stationarity of the probability density 
distributions of the synaptic weights $P(w_{ij})$
for various regimes. The distributions are stationary in the regimes of High and Low
Synchronization, namely when the Sisyphus Effect is absent. In these cases, they do not depend on the 
initial conditions and converge to the distribution reported in Fig.~\ref{fig:Pw} (a) and (b), 
corresponding to High and Low Synchronization, respectively. However, these distributions have 
some different features: in the HS regime $P(w_{ij})$ is essentially symmetric
and  peaked around $w_{ij}=1$; in the LS regime $P(w_{ij})$ is a skewed 
distribution peaked at $w_{ij} \simeq 0.5$ and with a tail extending towards larger values.
These results are consistent with the findings reported in~\cite{rossum2000}, where
the authors have shown that uncorrelated inputs lead to an unimodal distribution with a positive
skew (similar to the one reported in Fig.~\ref{fig:Pw} (b)), consistently with 
experimental findings~\cite{turrigiano1998,brien1998}.
Furthermore, in ~\cite{rossum2000} it has been also shown that
correlation among the inputs lead to a potentiation of the synapses
and to a more symmetric distribution. This is also the case for our model:
the distribution becomes more symmetric and peaked at a larger $w_{ij}$-value
when passing from the LS regime to the HS, characterized by a larger degree of
correlation among the neurons (see  Fig.~\ref{fig:Pw} (a,b)).

The situation changes completely in presence of the Sisyphus Effect,
since in this case the level of synchronization (of correlation) changes
continuously in time leading to a non-stationary distribution $P(w_{ij})$.
In particular, in Fig.~\ref{fig:Pw} (c,d) we reported the distributions estimated when
the system is almost completely desynchronized (synchronized) corresponding to $R \simeq 0.1$ 
($R \simeq 0.9$) shown in Fig.~\ref{fig:Pw} (c) (Fig.~\ref{fig:Pw} (d)).
These distributions have been obtained by averaging over $20$ different configurations 
of the system characterized by the desired level of synchronization.
It is evident that also in this case the synchronization favours the potentiation
of the synapses leading to the emergence of a positive tail extending towards $w_M$
(see Fig.~\ref{fig:Pw} (d)). 
However, in both situations two peaks are present in the distributions at $w_{ij} \simeq 1$
and $w_{ij} <  0.5$, indicating the coexistence of two subpopulations in the system
resembling the ones found in the HS and LS regime shown in  Fig.~\ref{fig:Pw} (a,b).
This result suggests that the Sisyphus Effect occurs on time scales which are
short with respect to the ones required by the synapses to relax towards an unimodal 
distribution. The distribution of the weights evolve in time recursively switching from
the distribution reported in panel (d) of Fig.~\ref{fig:Pw} to the one in panel (c)
and so on, for ever. It is important to stress that $P(w_{ij})$ does not tend to split in 
two groups peaked at $w_{ij}=0$ and $w_{ij}=w_M$, as it would occur in the models of STDP 
where potentiation and depression modify the synaptic weights by a fixed amount, 
irrespective of the actual value of $w_{ij}$~\cite{rossum2000}.

\section{Physical Units}

The model introduced in the paper contained only adimensional units,
since these are more convenient to perform numerical simulations. However, the
evolution equation for the membrane potential (\ref{eq:V1})
can be easily re-expressed in terms of dimensional variables as follows
\begin{equation}
\label{eq:dim}
\tau_m \dot{\cal V}_{j}(\tilde t)= {\cal R}_{in} {\cal I}^b - {\cal V}_{j}(\tilde t)+ 
\tau_m {\cal G} {\cal E}_j( \tilde t) \, \quad\quad j=1,\cdots, N
\\ \quad ;
\end{equation}
where $\tau_m = 40$ ms is the membrane time constant (as reported in \cite{derchansky2008transition} for 
pyramidal cells in brain area CA1 of hyppocampus), ${\cal R}_{in}$ is the membrane
resistance, ${\cal I}^b$ represents the neural excitability, due to contributions
from neurons lying outside the local network and projecting on them. Furthermore,
${\tilde t} = t \cdot \tau_m$, the inverse pulse-width is ${\tilde \alpha} = \alpha/\tau_m$,
the field ${\cal E}_j = E_j / \tau_m$ has the dimensionality of a frequency and ${\cal G}$ of a potential.

For the other parameters/variables the transformation to physical units is simply given by
\begin{eqnarray}
{\cal V}_{j} &=& {\cal V}_r + ({\cal V}_{th} - {\cal V}_{r}) V_j\\
{\cal R}_{in} {\cal I}^b &=& {\cal V}_r + ({\cal V}_{th} - {\cal V}_{r}) a\\
{\cal G} &=& ({\cal V}_{th} - {\cal V}_{r}) g 
\end{eqnarray}
where ${\cal V}_{r}= -60$ mV, ${\cal V}_{th}=-50$ mV ~\cite{Sterratt2011Principles, Dayan2001Theoretical}.
Typical values of the parameters employed in this paper were $a=1.3$, $g=0.4$, $\alpha=9$
and they correspond to ${\cal R}_{in} {\cal I}^b =-47$ mV, ${\cal G}=4$ mV, 
$\tilde \alpha = 225$ Hz. For these choices of parameters the average firing rate of the single neurons 
in the non plastic network was $\simeq 29$ Hz, while it decreased to $\simeq 23$ Hz in presence
of STDP plasticity.
 
As far as the STDP parameters are concerned, from the data reported by Bi \& Poo 
in \cite{Bi1998Synaptic} it emerges that the synaptic strengths of 
hyppocampal glutamatergic neurons are potentiated (depressed) 
of $\simeq 110 \%$ ($\simeq 40\%$) by considering 60 consecutive pairs of pre- and 
post-synaptic spikes. This amounts to have potentiation (depression) amplitude
$p \simeq 0.016$ ($d \simeq 0.0066$) in the model employed to mimic STDP (see Eq.~(\ref{post})), 
therefore we can safely affirm that our choices were consistent with the experimental data. 
The widths of the learning time windows are $\tilde \tau_+ = 4$ ms and  $\tilde \tau_- = 12$ ms.
These values are comparable with the rise/decay time of the excitatory post-synaptic potentials
$1/\tilde \alpha =4.44$ ms, but definitely smaller than those measured in the experiments, 
namely $\tilde \tau_+ \simeq 13-19$ ms and $\tilde \tau_- \simeq 34$ ms \cite{Bi1998Synaptic,Froemke2002Spiketimingdependent}.

\end{document}